\documentclass[prd,aps,preprint,amsmath,nofootinbib,amssymb,eqsecnum,showkeys]{revtex4}
\usepackage{verbatim,graphics,graphicx,color,slashed}
\usepackage{ulem} 
\usepackage{hyperref}

\def\GeV{\textrm{GeV}}

\def\nn{\nonumber \\}

\def\lsim{\mathrel {\vcenter {\baselineskip 0pt \kern 0pt
    \hbox{$<$} \kern 0pt \hbox{$\sim$} }}}
\def\gsim{\mathrel {\vcenter {\baselineskip 0pt \kern 0pt
    \hbox{$>$} \kern 0pt \hbox{$\sim$} }}}
\def\slashchar#1{\setbox0=\hbox{$#1$}           
 \dimen0=\wd0                                 
  \setbox1=\hbox{/} \dimen1=\wd1               
\ifdim\dimen0>\dimen1                        
  \rlap{\hbox to \dimen0{\hfil/\hfil}}      
  #1                                        
  \else                                        
 \rlap{\hbox to \dimen1{\hfil$#1$\hfil}}   
   /                                         
  \fi}                                         %
\def\cpto{\mathrel {\vcenter {\baselineskip 0pt \kern 0pt
    \hbox{$CP$} \kern 0pt \hbox{$\longrightarrow$} }}}
\def\cptof{\mathrel {\vcenter {\baselineskip 0pt \kern 0pt
    \hbox{$~CP$} \kern 0pt \hbox{$\longleftrightarrow$} }}}

\begin{document}

\baselineskip=15pt

\preprint{}

\title{Interplay between new physics in one-loop Higgs couplings and the top-quark Yukawa coupling}

\author{Xiao-Gang He${}^{1,2,3}$\footnote{hexg@phys.ntu.edu.tw}}
\author{Yong Tang${}^{1}$\footnote{ytang@phys.cts.nthu.edu.tw}}
\author{German Valencia$^4$\footnote{valencia@iastate.edu}}
\affiliation{${}^{1}$Physics Division, National Center for Theoretical Sciences,\\ 
Department of Physics, National Tsing Hua University, Hsinchu, Taiwan}
\affiliation{${}^{2}$INPAC, Department of Physics, Shanghai Jiao Tong University, Shanghai, China}
\affiliation{${}^{3}$CTS, CASTS and Department of Physics, National Taiwan University, Taipei, Taiwan}
\affiliation{${}^{4}$Department of Physics, Iowa State University, Ames, IA 50011, United States}

\date{\today $\vphantom{\bigg|_{\bigg|}^|}$}

\date{\today}

\vskip 1cm
\begin{abstract}

After the discovery of a 126 GeV state at the LHC it is imperative to establish whether this particle really is the Higgs boson of the standard model. The early measurements have not yet pinpointed any of the Higgs couplings to fermions, the Yukawa couplings of the standard model. In this paper we study the values of the top-quark-Higgs coupling, $g_{ht\bar{t}}$, that are still allowed by the one-loop couplings of the Higgs to two gluons or two photons. We first assume that both the gluon fusion production of the Higgs and its decay into two photons proceed through loops with standard model particles only, albeit with an arbitrary top-Higgs coupling. We find that the current Higgs data still allows for 20\% deviations in $g_{ht\bar{t}}$ from its standard model value. We then investigate the effect of new particles contributing to the effective one-loop couplings. Specifically we consider a color octet electroweak doublet extension of the scalar sector and find that in this case $g_{ht\bar{t}}$ is allowed to deviate by 40\% from its standard model value by the current data.

\end{abstract}

\pacs{PACS numbers: }

\maketitle

\section{Introduction}

The two collaborations ATLAS and CMS at the Large Hadron Collider (LHC) have found a new resonant state of mass near 126 GeV \cite{Aad:2012tfa,Chatrchyan:2012ufa}, purported to be the standard model (SM) Higgs boson. 
The immediate task after the discovery of this  state  it  to establish  whether it really is the Higgs boson of the standard model or whether there is physics beyond it (BSM). In particular, the early measurements have not yet ascertained any of the Higgs couplings to fermions, the Yukawa couplings of the standard model. 

This leaves open the more general possibility of a symmetry breaking sector separate from a fermion mass generation sector as in the early technicolor models \cite{Farhi:1980xs}.  In general, even the scale of electroweak symmetry breaking and that of fermion mass generation need not coincide \cite{Appelquist:1987cf,Maltoni:2001dc} and this motivates us to investigate deviations in the top-quark-Higgs coupling from its SM value. Of course, this possibility is not as exotic as it may first appear as even a simple two Higgs doublet model relaxes the proportionality between the fermion mass and its coupling to the (lightest) Higgs boson.

We will not concern ourselves with a detailed model of this kind. Instead we will be interested in the purely phenomenological question of constraining the values of the top-quark coupling to the observed Higgs boson using the current data. The main ingredients for this study are the one-loop couplings of the Higgs boson, namely its production via gluon fusion and its decay into two photons and into $Z\gamma$ as  these modes involve a top-quark loop and therefore the top-quark Higgs coupling. The experimental uncertainty in these measurements allows for a range of possible top-quark Yukawa couplings which we will discuss first.

We then study the interplay between the allowed  range for the top-quark-Higgs coupling and BSM particles that can change the one-loop Higgs processes.  In particular we will consider a simple extension of the scalar sector of the SM with new scalars $S$ transforming as $(8,2,1/2)$ under the  SM gauge group $SU(3)_C\times SU(2)_L\times U(1)_Y$. This color octet, electroweak doublet, scalar extension of the SM is that of Ref.~\cite{Manohar:2006ga}, motivated by the requirement of minimal flavor violation \cite{Chivukula:1987py,D'Ambrosio:2002ex}, and we have recently constrained the relevant parameters with unitarity and vacuum stability arguments \cite{He:2013tla}.

\section{Modified top-quark-Higgs coupling}

In the SM the top-quark coupling to the Higgs boson is uniquely determined by its mass as the Yukawa interaction reads:
\begin{eqnarray}
{\cal L}_{htt}&=& y_t \bar{q}\ t\  \tilde\phi +\ {\rm h.c.}
\label{lyuk}
\end{eqnarray}
where $q$ is the third generation SM quark doublet, $\phi$ is the scalar doublet, $\tilde\phi_i =\epsilon_{ij}\phi_j$ and the top-quark acquires a mass when electroweak symmetry is broken and the Higgs field develops a vacuum expectation value (vev) $\langle\phi\rangle=v/\sqrt{2}$, $v\approx 246$~GeV. Eq.~\ref{lyuk} then leads to the couplings
\begin{eqnarray}
{\cal L}_{htt}&=& \frac{y_t v}{\sqrt{2}} \bar{t}\ t \left(1+\frac{h}{v}\right).
\end{eqnarray}
The $ht\bar{t}$ coupling, $g_{ht\bar t}$, is thus fixed by the top-quark mass as $g_{ht\bar t}=y_t /\sqrt{2}=m_t/v$.

Beyond the SM, however, this no longer holds. In a model independent manner we can describe physics BSM with an effective Lagrangian that respects the symmetries of the SM. If we accept the 126~GeV state observed at LHC as a fundamental scalar, the appropriate effective Lagrangian for BSM physics with terms up to dimension six is that of Buchmuller and Wyler \cite{Buchmuller:1985jz,Grzadkowski:2010es}. One sees that already at dimension six there are terms in the Lagrangian modifying Eq.~\ref{lyuk}. For example, the term
\begin{eqnarray}
{\cal L}_{6}&=&\frac{g_{u\phi}}{\Lambda^2}(\phi^\dagger \phi) \bar{q}\  t\ \tilde\phi +\ {\rm h.c.}
\label{dim6}
\end{eqnarray}
suffices to spoil the proportionality between the top-quark mass and its coupling to the Higgs boson. In the presence of this term the $ht\bar{t}$ coupling and the top-quark mass are modified to 
\begin{eqnarray}
g_{ht\bar{t}} &=& \frac{y_t}{\sqrt{2}} + 3g_{u\phi}\frac{v^2}{2\sqrt{2}\Lambda^2} \nonumber \\
m_t &=&y_t\frac{v}{\sqrt{2}} + g_{u\phi}v\frac{v^2}{2\sqrt{2}\Lambda^2}
\end{eqnarray}
This distinction is of course model dependent but for our phenomenological study we will simply allow for an arbitrary $ht\bar{t}$ coupling parametrized by $r_t$ defined by 
\begin{eqnarray}
{\cal L}_{eff}=r_t\ \frac{m_t}{v} \ \bar{t} t h.
\end{eqnarray}
We thus have $g_{ht\bar{t}} = (g_{ht\bar{t}})_{SM}\ r_t$.

\section{Color-octet scalars} 

To alter the loop induced couplings of the Higgs boson, we will consider a model in which the scalar sector is augmented with a color octet, electroweak doublet in keeping with minimal flavor violation ~\cite{Manohar:2006ga}. The phenomenology of this model has been studied extensively \cite{Gresham:2007ri,Gerbush:2007fe,Burgess:2009wm,Carpenter:2011yj,Enkhbat:2011qz,Cacciapaglia:2012wb,He:2011ti,Dobrescu:2011aa,Bai:2011aa,Dorsner:2012pp,Kribs:2012kz,Arnold:2011ra,He:2011ws,Cao:2013wqa,He:2013tla} so we keep its discussion to a minimum. 
The inclusion of the new multiplet $S$ introduces several new, renormalizable, interaction terms to the Lagrangian. Because $S$ has non-trivial $SU(3)_C\times SU(2)_L\times U(1)_Y$ quantum numbers, it will have corresponding gauge interactions. In addition there will be new terms in the Yukawa couplings and in the Higgs potential that are consistent with minimal flavor violation. Only three of the nine new parameters will affect our discussion in this paper and they appear in the scalar potential of Ref.~\cite{Manohar:2006ga} as
\begin{eqnarray}
V&=&\lambda\left(H^{\dagger i}H_i-\frac{v^2}{2}\right)^2+2M_S^2\ {\rm Tr}S^{\dagger i}S_i +\lambda_1\ H^{\dagger i}H_i\  {\rm Tr}S^{\dagger j}S_j +\lambda_2\ H^{\dagger i}H_j\  {\rm Tr}S^{\dagger j}S_i \nonumber \\
&+&\left( \lambda_3\ H^{\dagger i}H^{\dagger j}\  {\rm Tr}S_ iS_j +{\rm h.c.}\right)+\cdots
\label{potential}
\end{eqnarray}
The first term is the same as the SM scalar potential and we use the conventional definition of $\lambda$ and of  $v\sim 246$~GeV. The traces are over the color indices and the $SU(2)$ indices $i,j$ are displayed explicitly. 

For our numerical discussion we will eliminate $\lambda_3$  by using the custodial symmetry relation $2\lambda_3=\lambda_2$. We will also restrict the ranges of $\lambda_1$  and $\lambda_2$ according to their unitarity and vacuum stability constraints recently derived. In particular we will use two conditions:
\begin{itemize}
\item the tree-level unitarity constraint as described in Ref.~\cite{He:2013tla} which can be summarized by
\begin{equation} 
\left| 2\lambda_1+\lambda_2\right| \lsim 18
\label{l12bound} 
\end{equation}
\item the renormalization group improved (RGI) unitarity constraint of the coupled equations for $\lambda$, $\lambda_1$ and $\lambda_2$ satisfied up to a large scale of $10^{10}$~GeV. This produces an allowed region shown in Figure~3 of Ref.~\cite{He:2013tla} which we import for this paper (the blue region in Figure~\ref{l1l2bounds}). Roughly, it can be thought of as the area limited by $-1.5 \lsim \lambda_1 \lsim 1.2$ and $- 1.7 \lsim \lambda_2 \lsim 1.2$. 
\end{itemize}

The masses of the scalars contributing to the one-loop Higgs couplings to gluons and photons are given in terms of the parameters of Eq.~\ref{potential} by \cite{Manohar:2006ga}
\begin{eqnarray}
m^2_{S^{\pm}} & = & m^2_S + \lambda_1 \frac{v^2}{4},\nonumber \\
m^2_{S^{0}_R}& = & m^2_S + \left(\lambda_1 + \lambda_2 + 2 \lambda_3 \right) \frac{v^2}{4},\\
m^2_{S^{0}_I}& = & m^2_S + \left(\lambda_1 + \lambda_2 - 2 \lambda_3 \right) \frac{v^2}{4},\nonumber
\end{eqnarray}
Their contribution to the effective couplings is reviewed in the Appendix.

\section{Numerical Results}

Fits to the LHC Higgs data already exist in the literature \cite{Carmi:2012yp,Azatov:2012bz,Espinosa:2012ir,Li:2012ku,Ellis:2012rx,Azatov:2012rd,Klute:2012pu,Azatov:2012wq,Low:2012rj,Corbett:2012dm,Buckley:2012em,Montull:2012ik,Espinosa:2012im,Carmi:2012in,Banerjee:2012xc,Bertolini:2012gu,Bonnet:2012nm,Plehn:2012iz,Espinosa:2012in,Moreau:2012da,Masso:2012eq,Corbett:2012ja,Belanger:2012gc,Cheung:2013bn,Cheung:2013kla,Falkowski:2013dza,Dumont:2013wma,Giardino:2013bma,Ellis:2013lra,Djouadi:2013qya,Chang:2013cia,Hayreter:2013kba} and we use Ref.~\cite{Giardino:2013bma} for our discussion. The relevant results from that reference are the fits to the one-loop effective couplings $r_\gamma$ and $r_g$ defined by
\begin{eqnarray}
{\cal L}_{hgg,h\gamma\gamma} &=& r_\gamma c_{SM}^{\gamma}\frac{\alpha}{\pi v} \ h F_{\mu\nu}F^{\mu\nu}+\ r_g c_{SM}^{g}\frac{\alpha_s}{12\pi v} \ h G^a_{\mu\nu}G^{a\mu\nu}
\end{eqnarray}
where the SM contributions $c_{SM}^{\gamma}$ and $c_{SM}^{g}$ are reviewed in the Appendix. 

\subsection{Color-octet scalars}

\begin{figure}[tbh]
\includegraphics[width=0.45\textwidth]{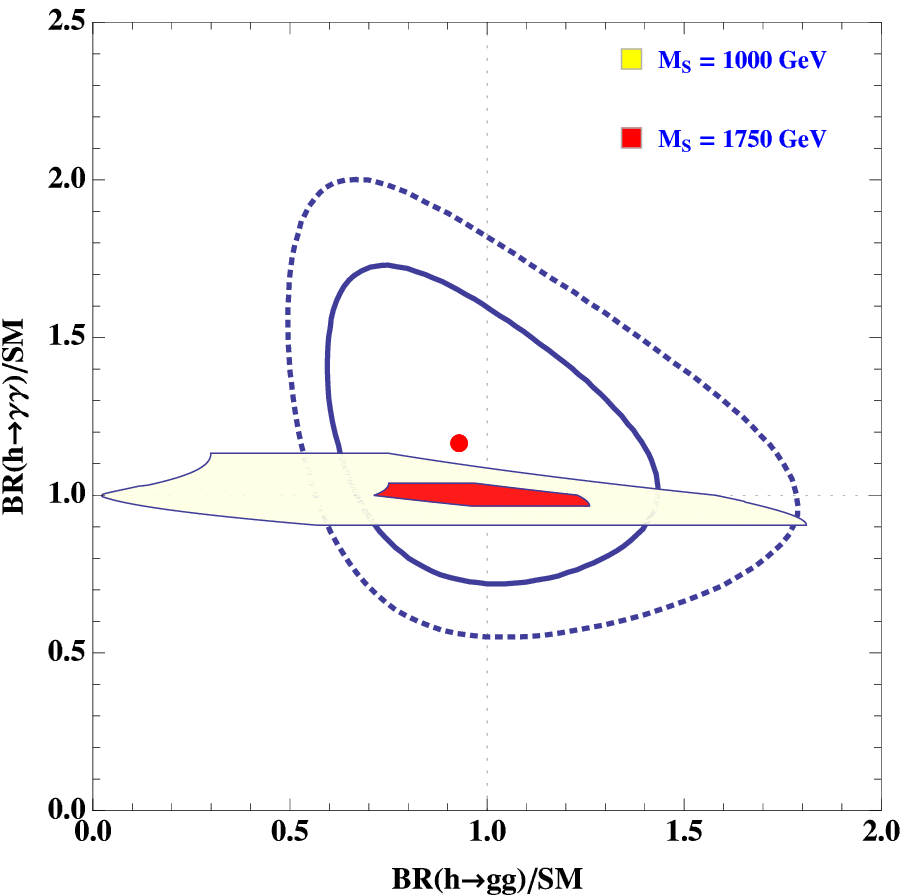}
\includegraphics[width=0.45\textwidth]{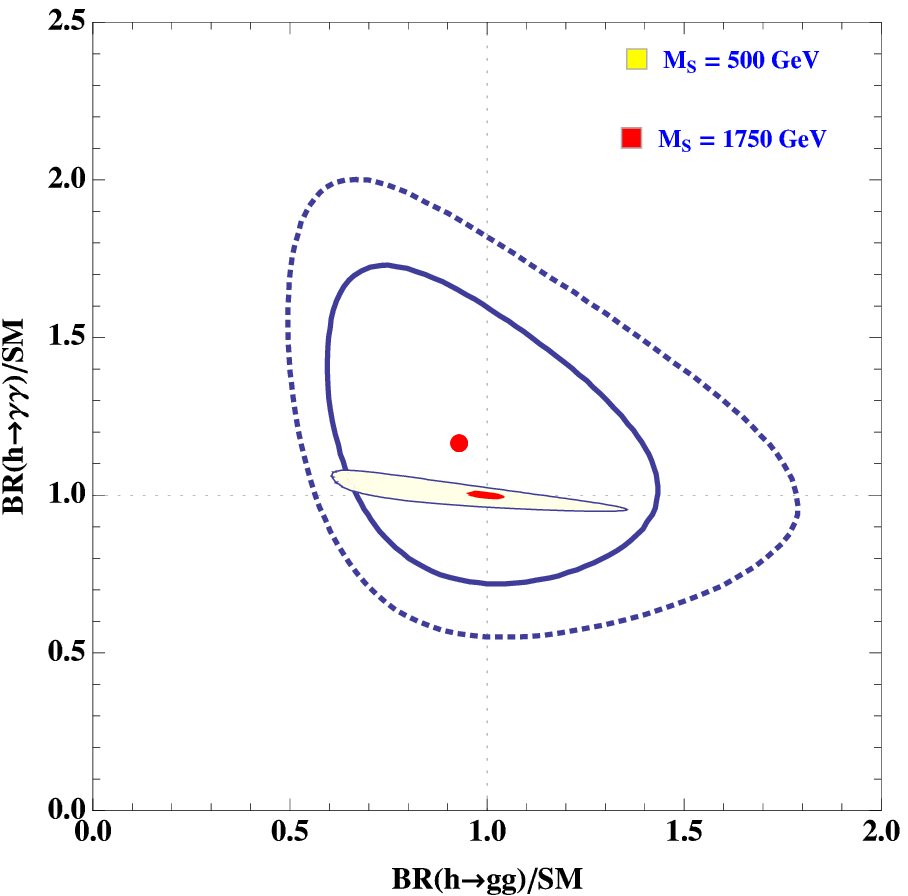}
\caption{Best fit to $BR(h\to \gamma\gamma)$ vs $BR(h\to gg)$ from Ref.~\cite{Giardino:2013bma}: the red dot is the best fit, the solid and dashed curves show the $1\sigma$ and $2\sigma$ allowed regions respectively. In the left panel we have superimposed the range of predictions in the color-octet model for two values of $M_S$ and values of $\lambda_{1,2}$ spanning the parameter space allowed by tree-level unitarity. In the right panel we span the parameter space allowed by the RGI unitarity conditions up to $10^{10}$~GeV.}
\label{octet}
\end{figure}

In Figure~\ref{octet} we show the results of the fit from  Ref.~\cite{Giardino:2013bma}, the red dot corresponding to the best fit to the data and the solid and dashed contours being the $1\sigma$ and $2\sigma$ regions respectively. The SM is the point $(1,1)$ in this plot as both axes are normalized to the SM rates. We have superimposed to this figure the results of adding a scalar color-octet. The figure on the left shows the regions obtained with parameters $\lambda_1$ and $\lambda_2$ that satisfy the tree-level unitarity constraint. We illustrate two cases, the yellow (larger) region for $M_S=1$~TeV and the red (smaller) region  for $M_S=1.75$~TeV. As expected, the region of possible rates shrinks as $M_S$ increases and approaches the SM point. The figure on the right is obtained with parameters $\lambda_1$ and $\lambda_2$ that satisfy the RGI unitarity conditions up to a scale of $10^{10}$~GeV. Again, we have illustrated two regions: a larger yellow one corresponding to $M_S=0.5$~TeV; and a smaller red one for $M_S=1.75$~TeV. We see that for values of $\lambda_1$ and $\lambda_2$ satisfying the RGI conditions up to a high scale, the corrections to the $h\to gg$ and $h\to \gamma\gamma$ rates are always within (or very close to) the $1\sigma$ fit to LHC data.

\begin{figure}[tbh]
\includegraphics[width=0.6\textwidth]{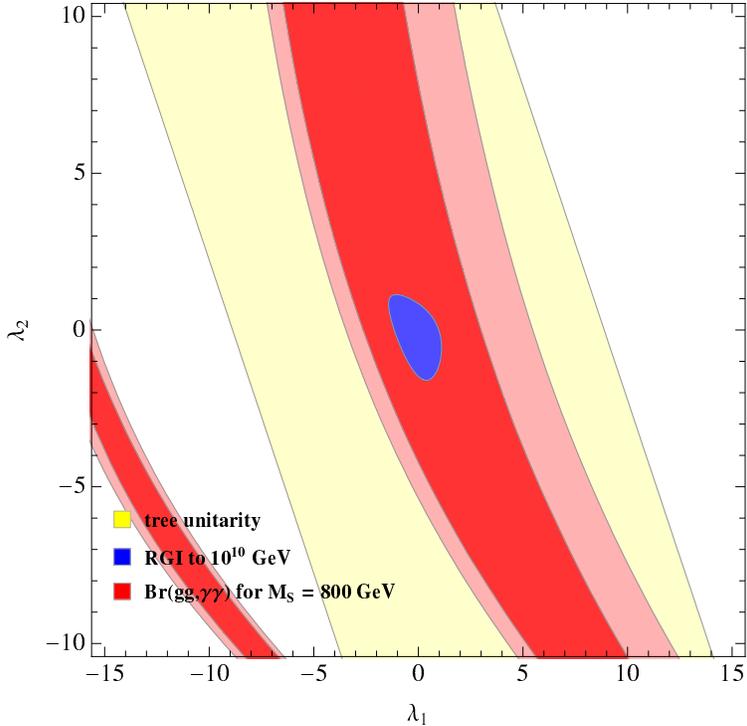}
\caption{Allowed $\lambda_1-\lambda_2$ parameter space from Ref.~\cite{He:2013tla} (yellow and blue regions as discussed in the text), superimposed with the regions allowed by the $BR(h\to \gamma\gamma)$ and $BR(h\to gg)$ at $1\sigma$ (dark red) and $2\sigma$ (light red). }
\label{l1l2bounds}
\end{figure}

We can turn the argument around and use the measured $BR(h\to \gamma\gamma)$ and $BR(h\to gg)$ to place additional constraints on the parameters $\lambda_1-\lambda_2$ of the color-octet scalar potential. We show this result in Figure~\ref{l1l2bounds}. This figure reproduces the allowed parameter space from tree-level unitarity (yellow) and RGI unitarity up to $10^{10}$~GeV (blue) from Ref.~\cite{He:2013tla}. On it we superimpose in dark (light) red the regions allowed at $1\sigma$ ($2\sigma$) by the $BR(h\to \gamma\gamma)$ and $BR(h\to gg)$ fit assuming the top-quark coupling is as in the SM. The two separate regions correspond to constructive (upper, larger regions) and destructive interference (lower narrow regions) with the SM respectively.

\subsection{Allowed range for $ht\bar{t}$ coupling}

\begin{figure}[tbh]
\includegraphics[width=0.75\textwidth]{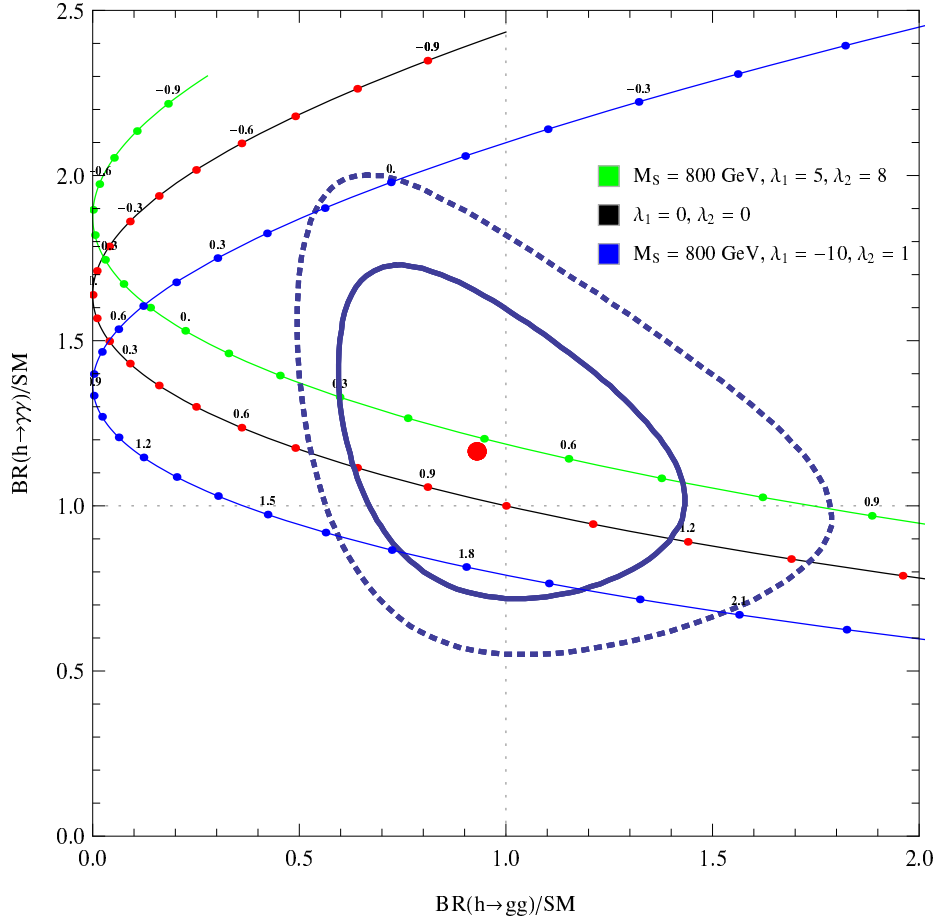}
\caption{$BR(h\to \gamma\gamma)$ vs $BR(h\to gg)$ as a function of $r_t$ (dots along the curves) for three cases. The parameters chosen for the black curve correspond to no additional scalars.}
\label{topx}
\end{figure}

We next consider the possibility of an arbitrary top-quark coupling to the Higgs boss, $r_t$, and illustrate three cases in Figure~\ref{topx}. The black curve shows the rates obtained by allowing $r_t$ to differ from 1 but without additional scalars (hence $\lambda_{1,2}=0$). We see that at the $1\sigma$ level, the data permits a 20\% excursion from the SM value, as the range $0.8 \lsim r_t \lsim 1.2$ is allowed. Notice that for $r_t =1$, which is the SM, the values for  $BR(h\to \gamma\gamma)$ and  $BR(h\to gg)$ are not the closest point to the best fit values to data. In the SM (no color octet), the 
amplitude for $h \to \gamma \gamma$ has contributions from a W loop and a top loop with different signs and with the latter being proportional to $r_t$. Allowing $ r_t$ to be smaller than 1, the cancellation between W and top-quark loops is reduced resulting in a larger branching ratio for $h \to \gamma\gamma$ and therefore in a better fit to the data. If one only considers this decay mode, $r_t \sim 0.6$ corresponds to the central value of the fit. However, varying $r_t$ will also modify $h \to gg$ whose amplitude is proportional to it. Taking both rates into account, the best fit is closer to $r_t \sim 0.95$.
  
New physics contributing to the loop amplitudes, such as the color-octet scalars, modifies the allowed $r_t$ range and we illustrate this with the green and blue curves in Figure~\ref{topx}. Values of $\lambda_{1,2}$ satisfying the RGI condition up to $10^{10}$~GeV result in minimal modifications, so we show two cases in which $\lambda_{1,2}$ are only required  to satisfy the tree-level unitarity bound.  A reversal of sign in $r_t$ is not allowed at the $2\sigma$ level.

\begin{figure}[tbh]
\includegraphics[width=0.32\textwidth]{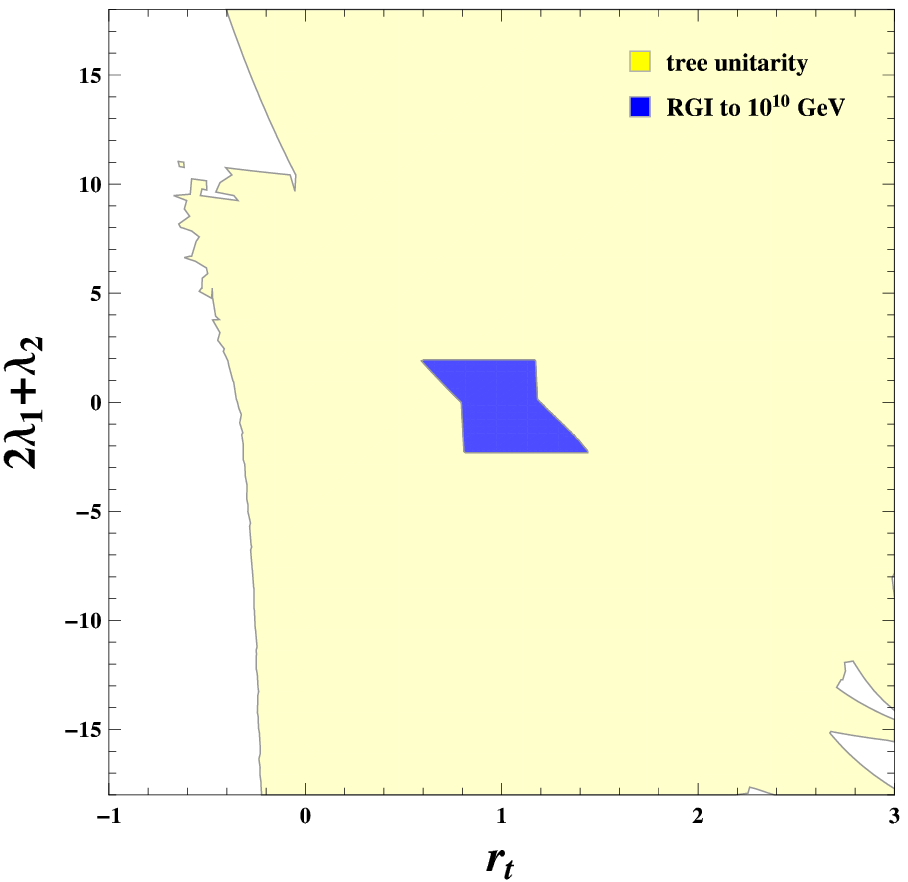}
\includegraphics[width=0.32\textwidth]{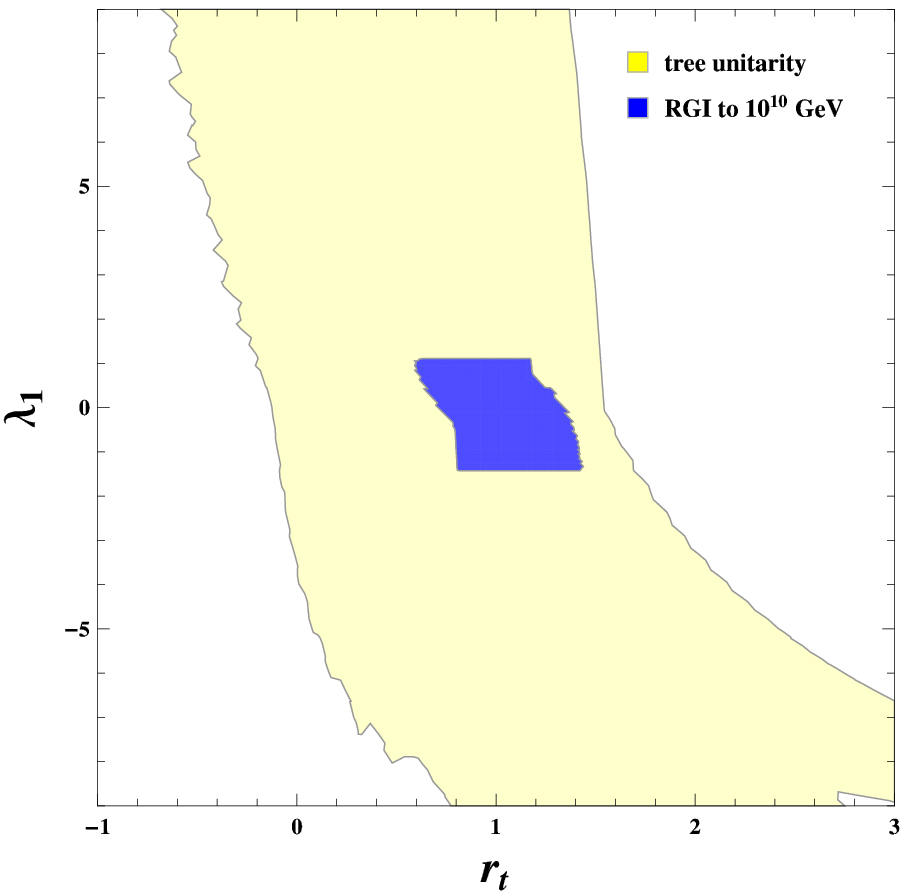}
\includegraphics[width=0.32\textwidth]{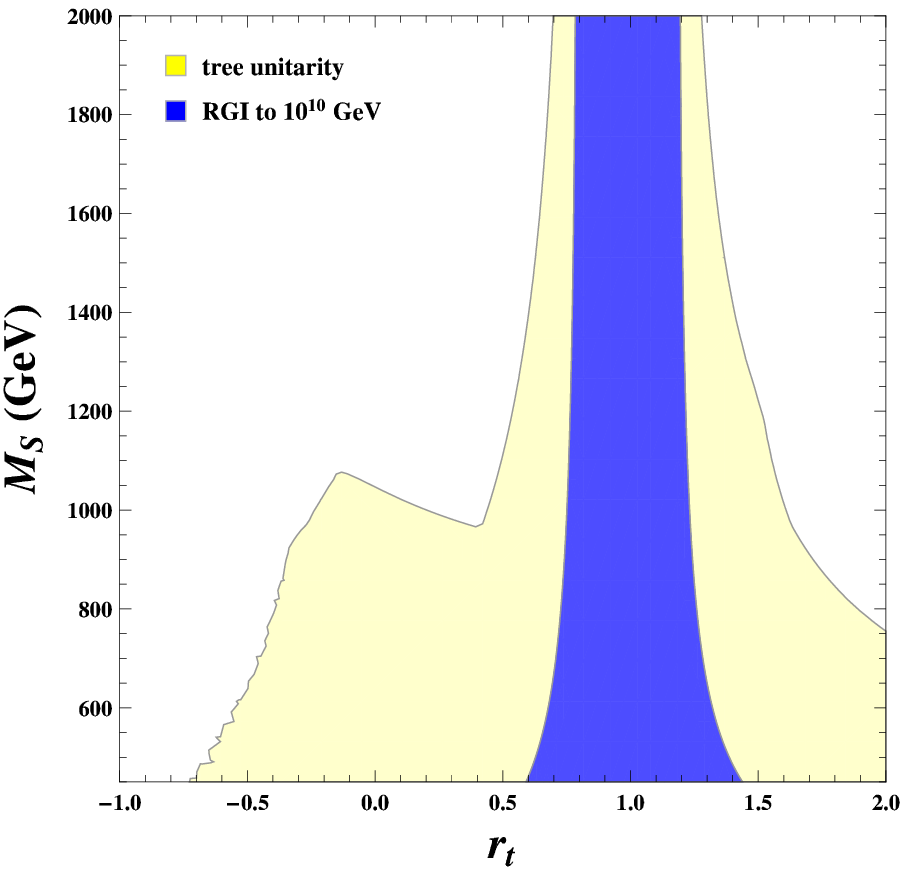}
\caption{Three views of the allowed parameter space at $1\sigma$. The larger region (yellow) allowed by tree level unitarity and the smaller region (blue) allowed by the RGI unitarity condition up to $10^{10}$~GeV. We display the allowed values of $r_t$ for ranges in $2\lambda_1+\lambda_2$ (left) and $\lambda_1$ (center) and $M_S$ (right).}
\label{otherviews}
\end{figure}

We collect in Figure~\ref{otherviews} three views of the allowed parameter space in the color-octet model as a function of $r_t$ such that the one-loop rates stay within the $1\sigma$ region of the fit.  The larger regions (yellow) span the $\lambda_1-\lambda_2$ parameter space allowed by tree level unitarity and the smaller regions (blue) span the $\lambda_1-\lambda_2$ parameter space allowed by the RGI unitarity condition up to $10^{10}$~GeV. We display the allowed values of $r_t$ for ranges in $M_S$, $2\lambda_1+\lambda_2$ and $\lambda_1$. The last two are chosen because $h\to gg$ depends mostly on $2\lambda_1+\lambda_2$ whereas $h\to \gamma\gamma$ depends mostly on $\lambda_1$. The plots illustrate how a deviation in $r_t$ from one can be compensated by the presence of additional scalars in the loop to end up with rates matching the observed ones. 

\section{Other modes}

In the previous discussion of constraints on the top-quark coupling to Higgs, we have implicitly assumed that changing the top Yukawa coupling does not modify the fitted contours from Ref.~\cite{Giardino:2013bma}. This assumption is justified for the following reasons. The first one is that the main channel for Higgs production is gluon-gluon fusion in which the top quark only appears in the loop. In the decay channel, only Higgs to di-photon is involved with top. So the effects of changing the top coupling can be parametrized as $Br(h\rightarrow gg)$ and $Br(h\rightarrow \gamma\gamma)$. The second reason is that the direct measurement of the top Yukawa through  $pp \rightarrow t\bar{t}h$ is not yet very restrictive, the current limit being about 5 times the SM value\cite{Chatrchyan:2013yea}. At present $r_t\in[-1,1]$ is  allowed by $pp \rightarrow t\bar{t} h$ alone. In addition, since the color-octet scalar does not develop a vev, the tree level $hWW$ and $hZZ$ couplings are not changed.

\begin{figure}[tbh]
\includegraphics[width=0.6\textwidth]{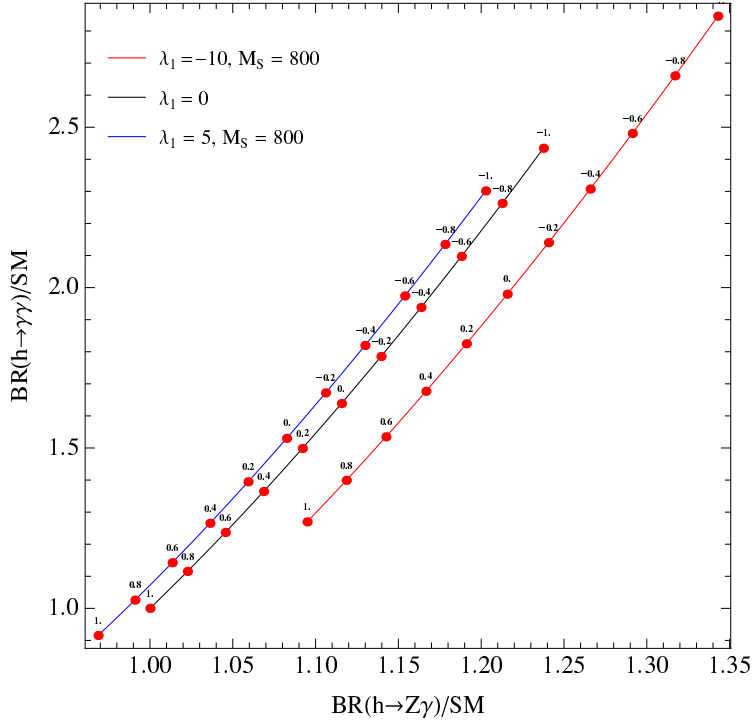}
\caption{Correlation between $Br(h\to Z\gamma)$ and $Br(h \to \gamma\gamma)$ for $M_S=800~\GeV$ and three values of $\lambda_1$ as a function of $r_t$ (indicated along the red dots).}
\label{Zgvsgg}
\end{figure}

The introduction of  a color-octet, however,  affects other loop induced processes such as $h \to Z \gamma$. We now study the predicted region allowed for this branching ratio. The detailed contributions are again reviewed in the Appendix. 
In general, the contribution from the new scalars to the decay $h\rightarrow \gamma\gamma$ is positively correlated with the contribution to $h\rightarrow Z\gamma$, as shown in Fig.~\ref{Zgvsgg} and both depend only on $\lambda_1$. We illustrate three different values of $\lambda_1$ allowed by tree-level unitarity and by $Br(h\rightarrow gg)$ and $Br(h\rightarrow \gamma\gamma)$ at $1\sigma$. In each case we also indicate the effect of varying $r_t$ in the range $r_t\in[-1,1]$. We see that $h \to Z \gamma$ is less sensitive to both the color-octet scalars and the variations of $r_t$ than 
$h \to \gamma\gamma$.

\section{summary}

The LHC has found a Higgs boson of mass near 126 GeV. The current available data have not yet pinpointed any of the Higgs couplings to fermions, the Yukawa couplings of the standard model. In this paper we have studied phenomenological  constraints on the values of the top-quark Yukawa coupling to the observed Higgs boson.  Data currently available to constrain this coupling  comes from the one-loop  Higgs boson amplitudes, namely its production via gluon fusion and its decay into two photons. The best fit value is away from the SM prediction although within the 1$\sigma$ region. At 1$\sigma$, $0.8 \lsim r_t \lsim 1.2$ is allowed by the data. 

We propose studying any deviation in terms of an interplay between new physics in the one-loop Higgs couplings and the top-quark Yukawa coupling. We have used a well motivated model, the color octet model, to illustrate the effect of new particles contributing to the loop amplitudes. The color octet effects on these decays are already severely constrained from unitarity considerations, but they can still play a role in these modes. In particular they relax the allowed range (at $1\sigma$) for the top-quark-Higgs coupling to  $0.6 \lsim r_t \lsim 1.4$. We pointed out that both the color-octet scalars and the variations in $r_t$ also play a role in $h \to Z \gamma$, although to a lesser extent.

\begin{acknowledgments}

The work of XGH and YT was supported in part by NSC of ROC, and XGH was also supported in part by NNSF(grant No:11175115) and Shanghai science and technology commission (grant No: 11DZ2260700) of PRC. The work of GV was supported in part by DOE under contract number DE-SC0009974. 

\end{acknowledgments}

\appendix

\section{Higgs Production and decay}

For completeness we review  the main ingredients in the (leading order) one-loop calculation of Higgs boson production in gluon fusion and its decay into two photons.  For a general discussion BSM we need to recall the different types of particles that can contribute to these two processes.

It is standard to parameterize the one-loop results with effective operators for  $hgg$ and $h\gamma\gamma$
\begin{equation}
\mathcal{L}_{\textrm{eff}}=c_g \frac{\alpha_s}{12\pi v}h G^a_{\mu\nu}G^{a\mu\nu}+c_\gamma \frac{\alpha}{\pi v}h F_{\mu\nu}F^{\mu\nu}.
\end{equation}
Different kinds of new particles such as a complex scalar  $S$, a Dirac fermion $f$, and a charged and colorless vector $V_\mu$ that couple to the Higgs as
\begin{equation}
\mathcal{L} = -  c_s \frac{2 M_S^2}{v}  h  S^\dagger S  - c_f \frac{M_f}{v} h \bar f f  +  c_V \frac{2 M_V^2 }{v} h V_\mu^\dagger  V^\mu.
\label{efftree}
\end{equation}
contribute to the effective Higgs coupling to gluons and to photons as \cite{Ellis:1975ap,Ioffe:1976sd,Shifman:1979eb,Djouadi:2005gi}
\begin{eqnarray}
\delta c_g   &=&  \frac{3C_2(r_s)}{ 2}  c_s A_s(\tau_s) +  \frac{3C_2(r_f)}{2} c_f A_f(\tau_f),\\
\delta c_\gamma    &=&   \frac{ N(r_s) Q_s^2 }{8}  c_s A_s(\tau_s) + \frac{ N(r_f) Q_f^2}{8} c_f A_f(\tau_f)  -  \frac{Q_V^2}{8} c_V A_V(\tau_V),
\end{eqnarray}
where $\delta c_i = c_i - c_{i,\rm SM} $,  $C_2(r)$ is the quadratic Casimir of the color representation $r$, and $N(r)$ is the number of colors of the representation $r$. The functions $A_i$ are defined as
\begin{eqnarray}
A_s(\tau) &\equiv& \frac{1}{\tau^2} \left [ f(\tau)  - \tau \right ]\,,
\nn
A_f(\tau) &\equiv& \frac{2}{\tau^2} \left [ (\tau-1)f(\tau)  + \tau \right ]\,,
\nonumber \\
A_V(\tau) &\equiv& \frac{1}{\tau^2}\left[3(2\tau-1)f(\tau)+3\tau+2\tau^2\right]\,,
\nonumber \\
f(\tau)
&\equiv&  \left\{ \begin{array}{lll}
{\rm arcsin}^2\sqrt{\tau} && \tau \le 1 \\ -\frac{1}{4}\left[\log\frac{\sqrt{\tau}+\sqrt{\tau-1}}{\sqrt{\tau}-\sqrt{\tau-1}}-i\pi\right]^2 && \tau > 1 \end{array}\right. 
\end{eqnarray}
with $\tau_i  = m_h^2/4M_i^2$. The feature of $f(\tau) \simeq \tau+\frac{\tau^2}{3}$ when $\tau \simeq 0$, leads to $A_s(0)=\frac{1}{3}$, $A_f(0)=\frac{4}{3}$ and $A_V(0)=7$. 

In the physics BSM that we discuss in this paper the only additional particles are scalars with $C_2(r_s)=3$. The SM contribution through the top-quark has $N(r_f)=3$ and octet has $N(r_s)=8$. Deviations from the SM in the top-quark coupling are parameterized by $r_t= c_t$ in Eq.~\ref{efftree} above. Since the masses of the color octet scalars are not entirely due to the Higgs vev, $c_s$ are the ratio of the $v^2$-dependent mass term in $M^2_s$ to $M^2_s$, for instance, $c_{S^{\pm}}=(\lambda_1 v^2)/(4\ m^2_{S^\pm})$.

In general, $hZ\gamma$ is also modified. Parametrized as $c_{Z\gamma} \frac{\alpha}{\pi v}h Z_{\mu\nu}F^{\mu\nu}$, we have \cite{Cahn:1978nz,Bergstrom:1985hp,Gunion:1989we,Djouadi:1996yq,Chen:2013vi}
\begin{align}
 c_{Z\gamma}=\frac{1}{8}\bigg{[}& c_VQ^2_V\cot\theta A_1^{Z\gamma}(\frac{1}{\tau_V},\frac{1}{\lambda_V})
+ c_f N(r_f) (2Q_f\cdot g_{Z\bar{f}f})A_{1/2}^{Z\gamma}(\frac{1}{\tau_f},\frac{1}{\lambda_f}) \nonumber\\
& - c_s N(r_s)(2Q_S\cdot g_{ZSS})A_0^{Z\gamma}(\frac{1}{\tau_S},\frac{1}{\lambda_S})\bigg{]},
\end{align}
where $\lambda_i=\frac{M_Z^2}{4M_i^2}$.
In standard Model, top quark has $g_{Z\bar{f}f}=\frac{1-4Q_t \sin^2{\theta_W}}{2\sin{\theta} \cos{\theta_W}}$. And a octet scalar gives $g_{ZSS}=\frac{1-2\sin^2{\theta_W}}{2\sin{\theta_W} \cos{\theta_W}}$ . Here $\theta_W$ is the Weinberg angle.
The functions are defined as follows:
\begin{subequations}
\begin{align}
A_1^{Z\gamma}(x,y) &= 4(3-\tan^2\theta)I_2(x,y)+[(1+2x^{-1})\tan^2 \theta-(5+2x^{-1})]I_1(x,y),\\
A_{1/2}^{Z\gamma} (x,y) &= I_1(x,y)-I_2(x,y),\\
A_0^{Z\gamma}(x,y) &= I_1(x,y),
\end{align}
\end{subequations}
where
\begin{subequations}
\begin{align}
I_1(x,y) &= \frac{x y}{2(x-y)}+\frac{x^2 y^2}{2(x-y)^2}[f(x)-f(y)]+\frac{x^2 y}{(x-y)^2}[g(x)-g(y)],\\
I_2(x,y) &= -\frac{x y}{2(x-y)}[f(x)-f(y)].
\end{align}
\end{subequations}
For $x>1$ we have
\begin{subequations}
\begin{align}
f(x) &= \arcsin^2\sqrt{1/x},\\
g(x) &= \sqrt{x-1}\arcsin\sqrt{1/x}.
\end{align}
\end{subequations}
For small changes $Br(Z\gamma)$ and $Br(\gamma\gamma)$ are linearly correlated as
\[
\cfrac{Br(Z\gamma)/SM-1}{Br(\gamma\gamma)/SM-1}=\left(\frac{\cos^2{\theta_W}-\sin^2{\theta_W}}{\cos{\theta_W} \sin{\theta_W}}\right)\frac{A_0^{Z\gamma}(\frac{1}{\tau_S},\frac{1}{\lambda_S})}{A_0^{\gamma\gamma}(\tau_S)}\sqrt{\frac{Br_{SM}(\gamma\gamma)}{Br_{SM}(Z\gamma)}}.
\]

\end{document}